\documentclass[11pt,twoside]{article}

\usepackage{asp2004}
\usepackage{epsf}
\usepackage{psfig}
\usepackage{lscape}
\usepackage{graphicx}

\begin{document}
\title{\emph{Spitzer} IRS spectra of Luminous 8~$\mu$m Sources in the Large
  Magellanic Cloud}

\author{Catherine L.\ Buchanan\altaffilmark{1}, Joel H.\
Kastner\altaffilmark{1}, William J.\ Forrest\altaffilmark{2}, Bruce
J.\ Hrivnak\altaffilmark{3}, Raghvendra Sahai\altaffilmark{4}, Michael
Egan\altaffilmark{5}, Adam
Frank\altaffilmark{2}, \& Cecilia Barnbaum\altaffilmark{6}}
\altaffiltext{1}{Center for Imaging Science, Rochester Institute of
Technology, 54 Lomb Memorial Drive, Rochester NY 14623. Email:
clbsps,jhk@cis.rit.edu}
\altaffiltext{2}{Department of Physics \& Astronomy, University of
Rochester, Bausch \& Lomb Hall, P.O. Box 270171, Rochester, NY
14627-0171}
\altaffiltext{3}{Dept. of Physics and Astronomy, Valparaiso
University, Valparaiso, IN 46383}
\altaffiltext{4}{NASA/JPL, 4800 Oak Grove Drive, Pasadena, CA 91109}
\altaffiltext{5}{Air Force Research Laboratory; OASD (NII) Space
Programs, Suite 7000, 1851 S. Bell St., Arlington, VA 22202}
\altaffiltext{6}{Valdosta University, 1500 N Patterson Street,
Valdosta, GA 31698}

\begin{abstract} 
We have produced an atlas of Spitzer Infrared Spectrograph (IRS) spectra of
mass-losing, evolved stars in the Large Magellanic Cloud.  These stars were
selected to have high mass-loss rates and so contribute significantly to the
return of processed materials to the ISM.  Our high-quality spectra enable the
determination of the chemistry of the circumstellar envelope from the mid-IR
spectral features and continuum. We have classified the spectral types of the
stars and show that the spectral types separate clearly in infrared
color-color diagrams constructed from 2MASS data and synthetic IRAC/MIPS
fluxes derived from our IRS spectra. We present diagnostics to identify and
classify evolved stars in nearby galaxies with high confidence levels using
Spitzer and 2MASS photometry.  Comparison of the spectral classes determined
using IRS data with the IR types assigned based on NIR colors also revealed a
significant number of misclassifications and enabled us to refine the NIR
color criteria resulting in more accurate NIR color classifications of
dust-enshrouded objects.
\end{abstract}
\keywords{atlases --- stars: AGB and post-AGB --- Magellanic Clouds ---
infrared: stars ---  stars: mass loss --- circumstellar matter}
\section{Introduction} 

Asymptotic Giant Branch (AGB) stars dominate the return of processed materials
to the interstellar medium (ISM) and therefore are an important component of
galaxy evolution.  High mass-loss rate objects are heavily obscured, so they
can be missed in optical surveys and need to be observed in the infrared
(IR). We have conducted a study using the \emph{Spitzer Space
Telescope}
InfraRed Spectrograph (IRS; \citealt{hou04}) of
a sample of luminous 8~$\mu$m sources in the Large Magellanic Cloud (LMC). Our
sample was selected from a compilation of \emph{2MASS/MSX} sources
\citep{ega01}, using near-infrared color and magnitude criteria to target high
mass-loss, evolved objects.

\begin{figure}
\plotone{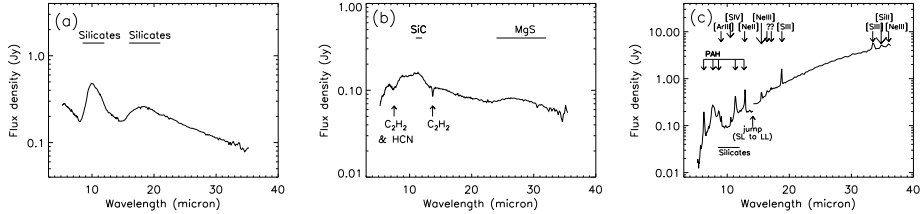}
\caption{{\it (a)} IRS spectrum of a typical O-rich object
  (MSX~LMC~264), showing broad silicate emission features at 10 and
  18~$\mu$m.  {\it (b)} Typical spectrum of a C-rich object
  (MSX~LMC~1400), showing broad SiC dust emission at 11.3~$\mu$m, and
  narrow C$_{2}$H$_{2}$ gas absorption at $\sim$7.5 and
  13.7~$\mu$m. {\it (c)} Typical spectrum of an emission-line object
  (MSX~LMC~22), showing narrow forbidden emission lines, broad PAH
  emission features, and red continuum indicative of cool
  dust. \label{fig:egspec}}
\end{figure}

IRS low-resolution spectra of 62 targets were obtained in Cycle 1. An atlas of
the spectra has been presented in \citet{buc06}. The spectra reveal continuum
and spectral features that allow the dominant chemistry of the dust
circumstellar envelope to be determined. Figure \ref{fig:egspec} shows typical
spectra of Oxygen-rich (O-rich), Carbon-rich (C-rich), and emission-line
objects.

\section{Results: Classification of IR-luminous Objects}   

\begin{figure}[!ht]
\plotfiddle{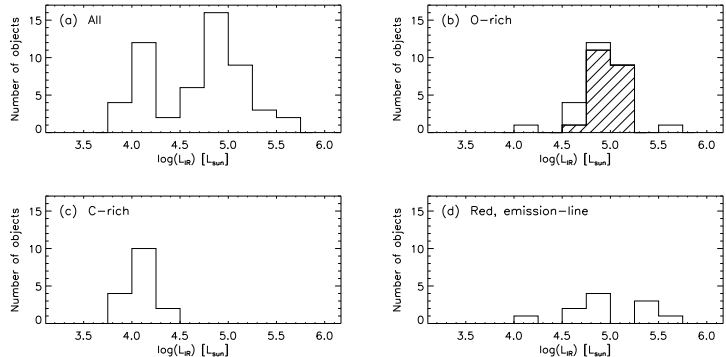}{1.8 in}{0}{50.0}{50.0}{-140}{0}
\protect\vspace*{-0.2 in}
\caption{Histograms showing the distributions of the IR luminosity for {\it
  (a)} the whole sample, {\it (b)} the O-rich stars, {\it (c)} the C-rich
  stars, and {\it (d)} the PAH-rich objects separately. The hatched regions of
  the histogram for O-rich objects indicate stars classified as RSGs based on
  their IR luminosities. The highest luminosity O-rich star that is not an RSG
  is an OH/IR supergiant.
  \label{fig:lumhist}}
\end{figure}

We find that almost all the O-rich stars are red supergiants (RSGs) with
infrared luminosities L$_{\rm{IR}} > 5 \times 10^{4}$~L$_{\odot}$ (Figure
\ref{fig:lumhist}).  The lack of lower luminosity O-rich AGB stars suggests
that massive supergiants do not have long enough He-burning lifetimes to
produce C-rich surfaces (and hence C-rich ejecta), despite the low metallicity
of the LMC.  The C-rich stars are all AGB stars with L$_{\rm{IR}} < 5 \times
10^{4}$~L$_{\odot}$. The large number of C-rich AGB stars stands in stark
contrast to the lack of O-rich AGB stars found. This result is consistent with
the hypothesis that C-rich stars form more easily than O-rich ones due to the
low metallicity environment of the LMC.  We find that the emission-line
objects are all H\,{\sc II} regions with L$_{\rm{IR}} >
10^{4}$~L$_{\odot}$. These objects were all expected to be Planetary Nebulae
(PNe) based on \emph{2MASS/MSX} colors \citep{ega01}, but optical and IR
images reveal diffuse nebulae around the objects, and the IRS spectra show a
jump in flux density between the Short-Low (5.2 -- 14~$\mu$m, slit width
3.6$^{\prime\prime}$) and Long-Low (14.0 -- 38~$\mu$m, slit width
10.5$^{\prime\prime}$) modules indicating the emission is extended on parsec
scales. The spectra also show a lack of high-ionization narrow emission lines
common in PNe (e.g., \citealt{kra02, ber04}).

\section{Diagnostics: Photometric Classifications} 

\begin{figure}
\plotfiddle{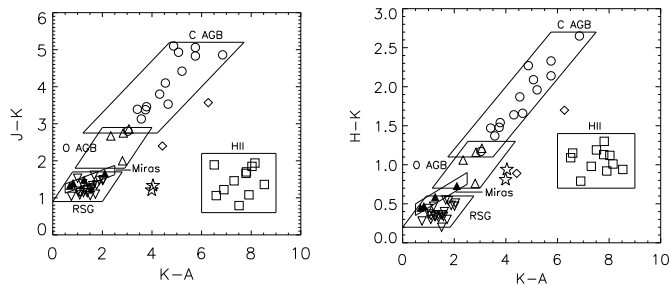}{1.6 in}{0}{50.0}{50.0}{-140}{0}
\protect\vspace*{-0.2 in}
\caption{\emph{2MASS/MSX} color-color diagrams. The symbols indicate the IRS
  spectral type: RSGs {\it (open, inverted triangles)}, O-rich AGB stars {\it
  (open triangles)}, Galactic Mira variables {\it (filled triangles)}, C-rich
  AGB stars {\it (open circles)}, H\,{\sc II} regions {\it (open squares)},
  B[e] supergiants (\citealt{kas06}; {\it open stars}), and OH/IR stars {\it
  (open diamonds)}. The boxes indicate the new NIR color criteria to classify
  IR-luminous LMC objects.
  \label{fig:nircols}} 
\end{figure}

Photometric tools such as color-color diagrams are invaluable for
identifying and classifying stellar objects. However, photometric
diagnostics must first be reliably associated with spectral properties
using spectroscopy. We used our IRS spectra to revise the \emph{2MASS/MSX}
classification criteria of \citet{ega01}.  In Figure \ref{fig:nircols}
we present \emph{2MASS/MSX} color-color diagrams showing our improved
photometric classification criteria.

\begin{figure}
\plotfiddle{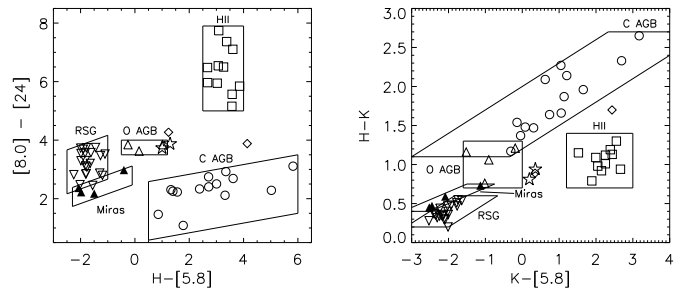}{1.6 in}{0}{50.0}{50.0}{-140}{0}
\protect\vspace*{-0.2 in}
\caption{Synthetic \emph{Spitzer/2MASS} color-color diagrams, showing
  our proposed criteria for classifying object types. Symbols
  are the same as Figure \ref{fig:nircols}.
  \label{fig:sstcols}} 
\end{figure}

One of the primary goals of our \emph{Spitzer} program was to provide a
spectroscopic basis for IRAC and MIPS photometric classifications of IR
sources in external galaxies. To that end, we derived synthetic IRAC and MIPS
magnitudes using the IRS spectra and the imaging filter spectral response
functions. Figure \ref{fig:sstcols} shows the resulting color-color diagrams
and diagnostics for identifying spectroscopic classes.


\section{Summary and Conclusions}

We have obtained low-resolution IRS spectra of a sample of 60 luminous
8~$\mu$m sources in the LMC and classified the sources according to their
spectral properties.  Almost all of the AGB stars in the sample are C-rich,
while the O-rich objects are luminous RSGs. We use our spectroscopic
classifications to develop revised infrared photometric diagnostics to
classify luminous IR sources, thus correcting inaccuracies in current
widely-used criteria.  We propose new photometric diagnostics to identify IR
sources in other galaxies, based on synthetic \emph{Spitzer} photometry and
existing \emph{2MASS/MSX} photometry. A full discussion of the data analysis
and additional results will appear in \citet{buc06}.

\acknowledgements 
Thanks to the conference organizers for a great meeting.  This work is based
on observations made with the \emph{Spitzer Space Telescope}, which is
operated by the Jet Propulsion Laboratory, California Institute of Technology
under a contract with NASA.  Support for this work was provided by NASA
through awards issued by JPL/Caltech. The IRS was a collaborative venture
between Cornell University and Ball Aerospace Corporation funded by NASA
through the Jet Propulsion Laboratory and Ames Research Center.

\end{document}